\documentclass[letterpaper,aps,prl,twocolumn,showcaps,superscriptaddress,showpacs,amsmath,amssymb]{revtex4}
\usepackage{graphicx}

\newcommand{\mean}[1]{\langle #1 \rangle}

\newcommand{\C}{\mathcal C}
\newcommand{\W}{\mathbb W}
\newcommand{\ld}{\psi}
\newcommand{\ldt}{\Psi}
\newcommand{\tobs}{t_\mathrm{obs}}
\newcommand{\sm}{s_\mathrm{m}}

\newcommand{\al}{\alpha}

\newcommand{\lam}{\lambda}
\newcommand{\kap}{\kappa}
\newcommand{\vhi}{\varphi}
\newcommand{\sig}{\sigma}
\newcommand{\sumij}{\sum_{\langle ij\rangle}}
\newcommand{\ad}{a^\dagger}
\newcommand{\ra}{\rightarrow}
\newcommand{\mf}{^{(\textrm{mf})}}

\begin{document}

\title{Space-time Phase Transitions in Driven Kinetically Constrained Lattice
  Models}

\author{Thomas Speck}
\affiliation{Department of Chemistry, University of California, Berkeley, CA
  94720, USA}
\affiliation{Chemical Sciences Division, Lawrence Berkeley National
  Laboratory, Berkeley, California 94720, USA}
\author{Juan P. Garrahan}
\affiliation{Department of Physics and Astronomy, University of Nottingham, 
  Nottingham, NG7 2RD, United Kingdom}

\begin{abstract}
  Kinetically constrained models (KCMs) have been used to study and understand
  the origin of glassy dynamics.  Despite having trivial thermodynamic
  properties, their dynamics slows down dramatically at low temperatures while
  displaying dynamical heterogeneity as seen in glass forming supercooled
  liquids.  This dynamics has its origin in an ergodic-nonergodic first-order
  phase transition between phases of distinct dynamical ``activity''.  This is
  a ``space-time'' transition as it corresponds to a singular change in
  ensembles of trajectories of the dynamics rather than ensembles of
  configurations.  Here we extend these ideas to driven glassy systems by
  considering KCMs driven into non-equilibrium steady states through
  non-conservative forces.  By classifying trajectories through their entropy
  production we prove that driven KCMs also display an analogous first-order
  space-time transition between dynamical phases of finite and vanishing
  entropy production. We also discuss how trajectories with rare values of
  entropy production can be realized as typical trajectories of a mapped
  system with modified forces.
\end{abstract}

\pacs{64.70.Pf, 05.40.-a}

\maketitle

%% ==== introduction ====

\textit{Introduction.--} The transformation of a supercooled liquid into an
amorphous solid glass at low temperatures has fascinated scientists for
decades~\cite{Reviews}, yet a comprehensive theory explaining the microscopic
origins of this ``glass transition" is still missing.  Several scenarios are
advocated in the literature.  These include: (i)~A transition to an ``ideal"
glass state reached at some critical value of a thermodynamic parameter such
as the temperature.  This transition may be thermodynamic, as in the random
first-order transition theory~\cite{RFOT}, or kinetic, as in mode-coupling
theory~\cite{MCT}. (ii)~An ``avoided'' glass transition, controlled by an
unreachable thermodynamic critical point~\cite{FLD}. (iii)~The glass
transition as a purely dynamical phenomenon that is not controlled by
structural changes and not related to a phase transition in a thermodynamic
sense. This approach~\cite{ARPC} considers dynamic heterogeneity~\cite{DH},
the observation of large spatial and temporal fluctuations in ``dynamical
activity'', to be the crucial feature of glass formers and derives most of its
predictions and insights from the study of kinetically constrained models
(KCMs) of glasses~\cite{KCM}.

Analytical studies of KCMs~\cite{garr07} and numerical investigations of
atomistic glass forming fluids~\cite{hedg09} have shown the existence of a
true phase transition in ensembles of trajectories. This ``space-time''
phase-transition is a first-order transition between two dynamical phases
which differ in their overall ``dynamical activity''~\cite{garr07,hedg09}.
This transition is not controlled by thermodynamic fields.  To reveal it one
needs to employ a statistical mechanics of trajectories whereby distributions
of trajectories are studied by controlling either dynamical activity or a
conjugate nonequilibrium field $s$~\cite{lecomte,garr07,hedg09,activity}.
Emergence of dynamic heterogeneity and other fluctuation phenomena can be seen
as manifestations of this underlying transition~\cite{ARPC}.

In this Letter we show the existence of analogous space-time phase transitions
in glassy systems driven to non-equilibrium stationary states by
non-conservative forces.  In this case the phase-transition is between
dynamical phases with finite and with vanishing entropy production.  We do so
by studying ensembles of trajectories via the large-deviation method in a
class of KCMs driven away from equilibrium by non-conservative forces. Such
forced KCMs~\cite{jack08,others} can be motivated experimentally by, for
example, pulling a tracer particle through a supercooled liquid or a colloidal
suspension~\cite{weeks}. The large-deviation method allows to study
sub-ensembles of trajectories with non-typical values of dynamical observables
in a way that sheds light onto the phase structure of the dynamics. We also
show that these sub-ensembles of rare trajectories can be generated as typical
trajectories by a mapping to an alternative system with modified forces. Such
a mapping might allow to bridge the conceptual gap between computer generated
biased ensembles and the experimental observation of space-time phase
transitions.

%% ==== entropy ====

\textit{Biased ensembles of trajectories.--} We first recall the results for
stationary equilibrium dynamics~\cite{garr07}. The activity
$K$~\cite{lecomte,garr07,hedg09,activity} is a dynamical (space-time) order
parameter. Information about the dynamics is encoded in the distribution of
$K$, or equivalently, in the dynamical partition sum
$Z(s)\equiv\mean{e^{-sK}}_0$. The brackets $\mean{\cdots}_0$ denote the
average over trajectories in the original ensemble, i.e., it is a sum over
trajectories $\C(t)$ weighted by $P_0[\C(t)]$. The field $s$ is conjugate to
the activity and the factor $e^{-sK}$ enhances the weight of trajectories with
either higher than typical activity ($s<0$), or with lower than typical
activity ($s>0$). One can think of $s$ as giving rise to a biased ensemble of
trajectories with weights $P_s[\C(t)]\propto P_0[\C(t)]e^{-sK[\C(t)]}$. We
refer to this biased ensemble as the $s$-ensemble~\cite{hedg09}.

Denoting $N$ the system size and $\tobs$ the length of trajectories, their
product $N\tobs$ is the ``volume'' in space-time. For large $N\tobs$, the
dynamical partition sum acquires a large deviation form~\cite{touchette} with
$Z(s;N\tobs)\sim e^{N\tobs\ld(s)}$. The large-deviation function $\ld(s)$
generates the moments of $K$ in the biased ensemble, e.g., the mean activity
rate is $\mean{K}/(N\tobs)=-\ld'(s)$ where $\mean{\cdots}$ now denotes the
average in the $s$-ensemble. The function $\ld(s)$ is akin to a
free-energy. For KCMs it has a first-order singularity at $s=0$ indicating a
space-time phase transition~\cite{garr07}. This transition is between an
ergodic phase of finite activity rate and a nonergodic phase of vanishing
activity rate. In the nonergodic phase the system, through the kinetic
constraints, is able to arrange trajectories such that the activity grows
sublinearly. These trajectories ``win'' at $s=0^+$ because active trajectories
with nonzero activity acquire a negative weight in the partition sum and thus
are exponentially suppressed for large $\tobs$. At $s=0^-$ the opposite occurs
and the dominant phase is the active, ergodic, one.

We now extend the $s$-ensemble approach to driven systems. Transitions between
configurations $\C$ occur with rates $w(\C\ra\C')$ implying the exit rate
$r(\C)\equiv\sum_{\C'}w(\C\ra\C')$. For a given trajectory $\C(t)$ of time
span $\tobs$ the system undergoes $K$ transitions $\C_{\al-1}\ra\C_\al$ at
times $t_\al$ where $\C_0$ is the initial state, i.e.,
$\C(t)\equiv(\C_0,\dots,\C_K)$. The entropy produced in the environment in a
single transition $\C\ra\C'$ is $\Delta
s(\C,\C')=\ln[w(\C\ra\C')/w(\C'\ra\C)]$. In analogy to the activity, we may
bias trajectories using the time-extensive medium entropy production
$\sm[\C(t)]\equiv\sum_{\al=1}^K\Delta s(\C_{\al-1},\C_\al)$, where we sum over
all configuration changes. The corresponding dynamical partition function is
\begin{equation}
  \label{eq:part:lam}
  Z(\lam;N\tobs) \equiv \mean{e^{-\lam\sm}}_0 \sim e^{N\tobs\ldt(\lam)}
\end{equation}
where $\lam$ is the parameter conjugate to $\sm$.  In analogy with the
activity, the mean entropy production rate in the $\lam$-ensemble is given by
$\mean{\sm}/(N\tobs)=-\ldt'(\lam)$. The partition sum (\ref{eq:part:lam}) has
a natural transfer matrix representation, and $\ldt(\lam)$ is given by the
largest eigenvalue of the operator~\cite{Lebowitz}
\begin{equation}
  \label{eq:lam}
  \W_\lam(\C',\C) = \left[\frac{w(\C'\ra\C)}{w(\C\ra\C')}\right]^\lam
  w(\C\ra\C') - r(\C)\delta_{\C\C'} .
\end{equation}
In what follows we study the dynamical phase structure of driven KCMs by
calculating their large-deviation functions $\ldt(\lam)$ from the
corresponding operators $\W_\lam$.

%% ==== FA model ====

\textit{Driven Fredrickson-Andersen model.--} The simplest KCM that displays
glassy features is the one-spin facilitated Frederickson-Andersen (FA)
model~\cite{fred84,KCM}.  We introduce a driven variant of the FA model with
evolution operator
\begin{equation*}
  % \label{eq:FA:driven}
  \W_0 \equiv
  \sumij\left\{ \left[c\ad_i+a_i-(c+\hat n_i)\right] \hat n_j
    + k_{i\ra j}(\ad_ja_i - \hat n_i) \right\}.
\end{equation*}
For simplicity we consider a bosonic version of the FA model, i.e. we allow
for multiple occupancy of sites~\cite{bosonic}. The sum runs over all pairs of
nearest neighbors. Here, $\ad_i$ and $a_i$ are creation and annihilation
operators at site $i$ with number operator $\hat n_i=\ad_ia_i$. The first term
describes the creation and annihilation of excitations at site $i$ with rate
$c\equiv e^{-\beta J}$, where $J$ is the energy scale and
$\beta\equiv1/(k_\mathrm{B}T)$ is the inverse temperature.  The dynamics is
constrained in the sense that at least one nearest neighbor must be excited to
allow transitions at site $i$. The second term allows for explicit diffusion
of excitations from site $i$ to one of its neighboring sites with rates
$k_{i\ra j}$~\cite{how}. Using Eq.~(\ref{eq:lam}), the time-evolution operator
for the $\lam$-ensemble reads
\begin{multline}
  \label{eq:FA:lam}
  \W_\lam = \sumij\left\{ \left[ \left(\frac{\hat n_i}{c}\right)^\lam c\ad_i +
      \left(\frac{c}{\hat n_i}\right)^\lam a_i - (c+\hat n_i)\right] \hat n_j
  \right. \\
  + \left. k_{i\ra j} \left[
      \left(\frac{k_{j\ra i}\hat n_j}{k_{i\ra j}\hat n_i}\right)^\lam
      \ad_ja_i - \hat n_i \right] \right\}.
\end{multline}

\begin{figure}[t]
  \centering
  \includegraphics[width=\linewidth]{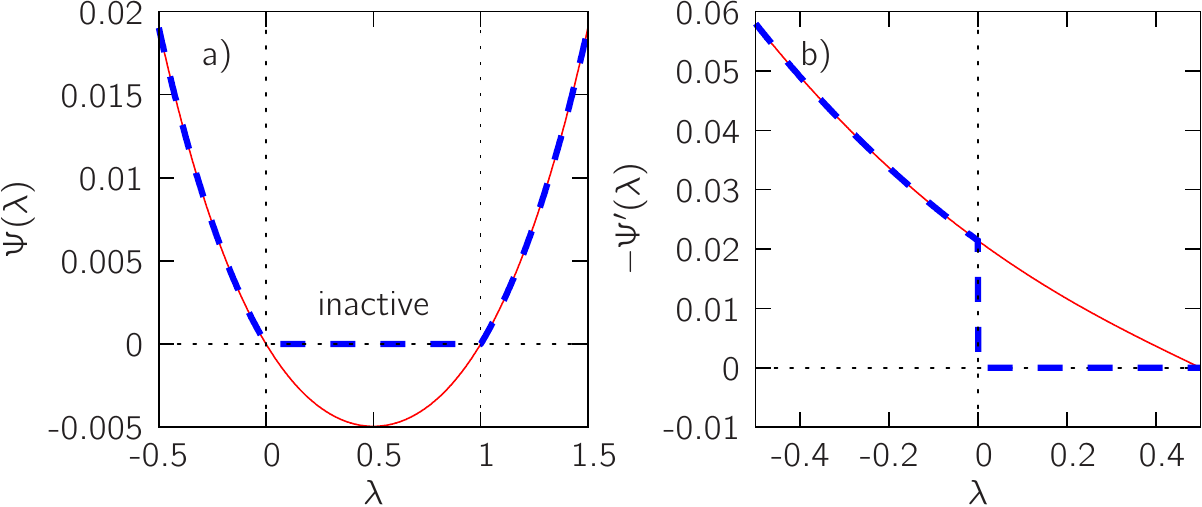}
  \caption{(a) Mean-field estimate of the large-deviation function
    $\ldt\mf(\lam)$ [Eq.~(\ref{eq:ldf:mf}), bold dashed] for the driven FA
    model with $c=0.2$, $k^-/k^+=0.7$.  There is a first-order transition from
    an active (entropy producing) dynamical phase to an inactive dynamical
    phase at $\lam=0$.  The ``reentrant'' transition at $\lam=1$ is due to the
    Gallavotti-Cohen symmetry. The thin solid line corresponds to the active
    branch for all values of $\lam$. (b) Mean entropy production rate
    $-\ldt'(\lambda)$.  }
  \label{fig:fa}
\end{figure}

We restrict our calculation to a one-dimensional lattice with $N$ sites and
periodic boundaries. Further, we consider spatially homogeneous rates: $k^+$
for a hop $i\ra i+1$ and $k^-$ for $i\ra i-1$ such that the steady state
solution for the mean density becomes independent of the site.  The largest
eigenvalue of (\ref{eq:FA:lam}) can be estimated in a mean-field approximation
by maximizing the function $W(\bar\phi,\phi)$ obtained from the (normally
ordered) operator~(\ref{eq:FA:lam}) by replacing $\ad\ra\bar\phi$ and
$a\ra\phi$.  There are two solutions to the Euler-Lagrange equations
$\partial W / \partial \phi=0$ and $\partial W / \partial \bar\phi=0$. In
terms of the mean density $n=\bar\phi\phi$ the two solutions read: $n=0$ and
$n=(c/16)(3+\sqrt{1+4\kap(\lam)/c})^2$.  They correspond to the inactive and
active phases, respectively. The $\lam$-dependence is contained in
\begin{equation}
  \label{eq:kap}
  \kap(\lam) \equiv \left(\frac{k^+}{k^-}\right)^\lam k^-
  +  \left(\frac{k^-}{k^+}\right)^\lam k^+ - (k^++k^-).
\end{equation}
The mean-field large deviation function becomes
\begin{equation}
  \label{eq:ldf:mf}
  \ldt\mf(\lam) =
  \begin{cases}
    0 & (0\leqslant\lam\leqslant1) \\
    2n^{3/2}(\sqrt n-\sqrt c) & (\lam\leqslant0~\text{and}~\lam\geqslant1).
  \end{cases}
\end{equation}
The function $\ldt\mf(\lam)$ and its first derivative are plotted in
Fig.~\ref{fig:fa}. The negative first derivative is the mean entropy
production rate. It shows a discontinuous transitions at $\lam=0$ between
finite entropy production ($\lam<0$, the active phase) and vanishing entropy
production ($\lam>0$, the inactive phase). For $\lam=1$, there is a second
``reentrant'' transition to an active phase now with a negative entropy
production, see Fig.~\ref{fig:fa}(a). This second transition is a reflection
of the transition at $\lam=0$ due to the Gallavotti-Cohen symmetry, which
enforces the relation $\ldt(1-\lam)=\ldt(\lam)$ on the large deviation
function~\cite{gall95,Lebowitz} through Eq.~\eqref{eq:kap}. Biasing with
$\lam>1$ corresponds to driving that essentially changes the sign of all
currents due to time-reversal symmetry.

%% ==== TLG ====

\begin{figure}[t]
  \centering
  \includegraphics[width=\linewidth]{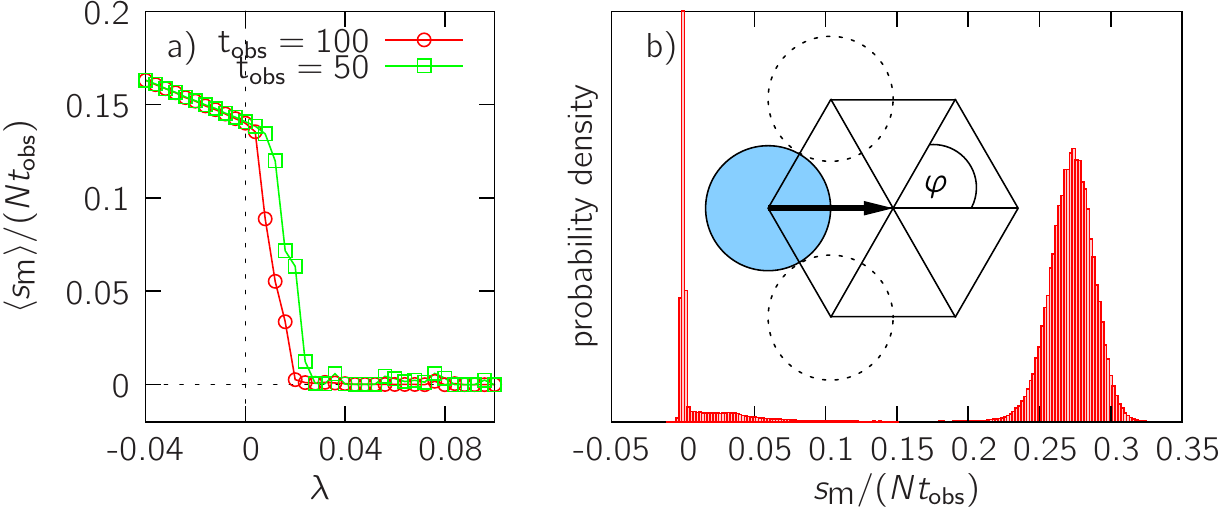}
  \caption{(a)~Numerical result for the mean entropy production rate in the
    (2)-TLG for parameters: $L=16$, $\rho=0.5$, and $f=0.5$. (b)~Bimodal
    distribution of the entropy production rate at $\lam=0.003$.  The inset
    shows the kinetic constraint of the (2)-TLG: the particle can only move
    along the arrow if the two indicated sites are vacant.}
  \label{fig:tlg}
\end{figure}

\textit{Driven constrained lattice gas.--} For a numerical study, we turn to a
driven variant~\cite{jack08,others} of a kinetically constrained triangular
lattice gas (TLG)~\cite{tlg,KCM}. Lattice gas particles diffuse on a
two-dimensional triangular lattice with periodic boundary conditions.  The
density of particles $\rho$ is conserved.  A particle can only move to a
neighboring site if the two adjacent sites are vacant, see inset of
Fig.~\ref{fig:tlg}(b).  This version of the model is called the
(2)-TLG~\cite{tlg}. In addition, we apply a force $f$ in the $x$-direction.
As a consequence the rates for allowed moves are $w(\vhi)=e^{(f/2)\Delta
  x_\vhi}$ where $\vhi=0,\pm\pi/3,\pm2\pi/3,\pi$ is the angle between the
displacement vector and the $x$-axis. In a single transition a particle moves
in the $x$-direction the distance $\Delta x_\vhi=2\cos\vhi=\pm1,\pm2$.  We
denote by $K_\vhi[\C(t)]$ the number of moves in the direction $\vhi$ in a
trajectory $\C(t)$.  The activity is then given by $K[\C(t)]=\sum_\vhi
K_\vhi[\C(t)]$, and the entropy production by $\sm[\C(t)]=f\sum_\vhi
K_\vhi[\C(t)] \Delta x_\vhi$.

We use transition path sampling~\cite{tps} to numerically generate biased
ensembles of trajectories, for details see
Ref.~\cite{hedg09}. Fig.~\ref{fig:tlg}(a) shows the mean entropy production
rate $\mean{\sm}/(N\tobs)$ per particle, from ensembles of $10^5$ trajectories
for $\lam<0$, and over $10^6$ trajectories for $\lam\geqslant 0$. There is
clear evidence of a transition between a phase of finite and a phase of
vanishing entropy production similar to Fig.~\ref{fig:fa}(b) for the driven FA
model.  Compared to the mean field solution the transition is smeared out and
shifted to $\lam\gtrsim0$ due to finite size effects. The data suggests that
the transition point moves towards $\lambda=0$ with increasing space-time
volume.  The probability density of the entropy production rate in
Fig.~\ref{fig:tlg}(b) for $\lam=0.003$ displays a bimodal character, as
expected from a first-order transition. Just like for the case of equilibrium
dynamics and the space-time transition in terms of activity~\cite{garr07}, we
expect the first-order transition between phases of distinct entropy
production to be generic in driven KCMs.

%% ==== mapping ====

\textit{Mapping ensembles.--} We now address the question of whether the biased 
ensembles of trajectories can be realized
experimentally.  In the $\lam$-ensemble, the weight of a trajectory segment
starting in state $\C$, surviving for $\Delta t$, and then jumping to state
$\C'$ reads
\begin{equation*}
  e^{-r(\C)\Delta t}w(\C\ra\C')e^{-\lam\Delta s(\C,\C')-N\Delta t\ldt(\lam)}.
\end{equation*}
We consider long trajectories and include the large deviation function
$\ldt(\lam)$ for normalization. Collecting terms containing $\Delta t$, the
exit rate $\tilde r$ in a modified dynamics must fulfill
\begin{equation}
  \label{eq:mod:exit}
  \tilde r(\C;\lam) = \sum_{\C'} \tilde w(\C\ra\C';\lam) = r(\C) + N\ldt(\lam),
\end{equation}
i.e., the difference between modified and original exit rate is given by
$\ldt(\lam)$ \textit{for all states}. Simply modifying the transition rates as
$we^{-\lam\Delta s}$ is not enough to give the exit rates $\tilde r$.
However, adding a sub-extensive term to $\sm$ does not change the large
deviation function. We therefore define the modified rates
\begin{equation}
  \label{eq:mod:rates}
  \tilde w(\C\ra\C';\lam) = w(\C\ra\C') e^{-\lam\Delta s(\C,\C')-\Delta
    u(\C,\C';\lam)}
\end{equation}
such that the sum of $\Delta u$ along single trajectories is sub-extensive in
$\tobs$. Due to antisymmetry $\Delta u(\C',\C)=-\Delta u(\C,\C')$ the sum
$\sum_\ell\Delta u=0$ vanishes along any closed loop $\ell$. This property
implies $\Omega-1$ independent quantities $\Delta u$, where $\Omega$ is the
number of configurations $\C$. The $\Omega$ equations~(\ref{eq:mod:exit}) then
determine $\{\ldt,\Delta u\}$. A similar result has been found in
Ref.~\cite{baul08} for biasing equilibrium towards a shear driven ensemble.

In the simplest case, $\Delta u(\C,\C';\lam)\equiv u(\C';\lam)-u(\C;\lam)$
with state function $u(\C;\lam)$ depending on $\lam$. The sum
$\sum_{\al=1}^K\Delta u(\C_{\al-1},\C_\al)=u(\C_K)-u(\C_0)$ is then a temporal
boundary term.  As we will see below, $u(\C)$ can be interpreted as an energy
function. The modified dynamics~(\ref{eq:mod:rates}) is still Markovian but
the rates become nonlocal since a transition from site $i$ can now depend on
the whole configuration $\C$. Even for simple KCMs configuration space is too
big to allow an explicit determination of $u(\C)$.

%% ---- illustration ----

\begin{figure}[t]
  \centering
  \includegraphics[width=\linewidth]{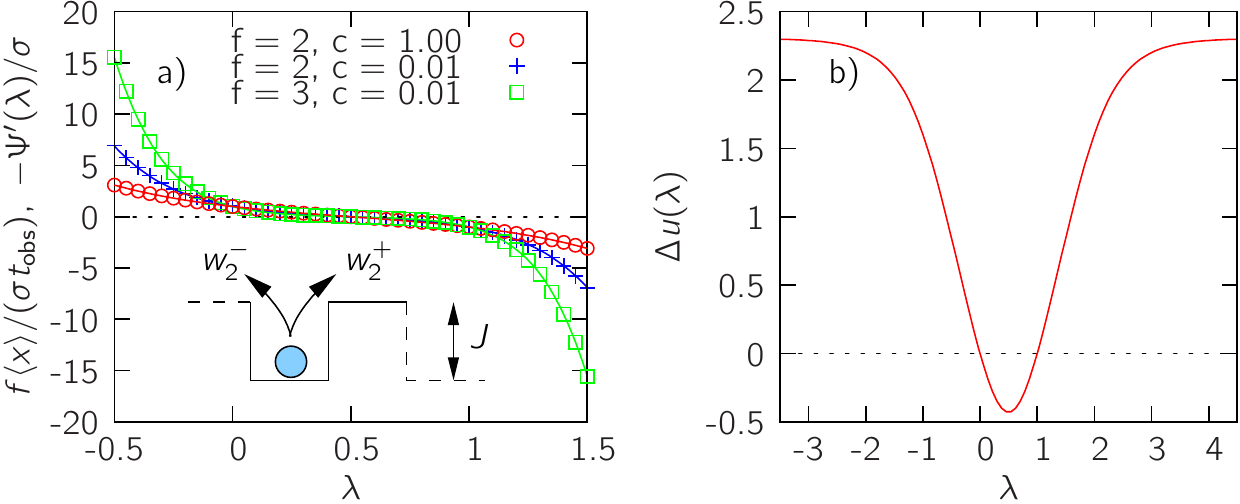}
  \caption{Particle moving in an infinite lattice with alternating energy
    levels (see inset). (a)~Symbols are numerical results of
    $f\mean{x}/(\sig\tobs)$ for a particle moving with rates
    Eq.~(\ref{eq:walk:rates}), where $\sig\equiv-\ldt'(0)$. The values agree
    with the normalized mean entropy production rate $-\ldt'(\lambda)/\sig$ in
    the $\lam$-ensemble (solid lines). (b)~The function $\Delta u(\lambda)$
    for $f=2$ and $c=0.01$.}
  \label{fig:walk}
\end{figure}

To illustrate the idea of mapping the $\lam$-ensemble to a physical dynamics
we consider a simple model related to the driven FA model, but without kinetic
constraints, that can be tackled analytically. A random walker moves in a
periodic lattice with alternating site energies, see inset of
Fig.~\ref{fig:walk}(a). The equilibrium rates are $w^\pm_1=1$ and $w^\pm_2=c$
for steps to the right (+) or left (-), where $c\equiv e^{-\beta J}$ and $J$
is the energy difference. Applying a driving force $f$, the entropy produced
in a single step is $\pm f$.  The total entropy production is $\sm=fx$, where
$x$ is the distance the particle has traveled. The modified rates from
Eq.~(\ref{eq:mod:rates}) are
\begin{equation}
  \label{eq:walk:rates}
  \begin{split}
    \tilde w^+_1 = e^{+f/2-\lam f-\Delta u}, \quad &
    \tilde w^+_2 = ce^{+f/2-\lam f+\Delta u}, \\
    \tilde w^-_1 = e^{-f/2+\lam f-\Delta u}, \quad &
    \tilde w^-_2 = ce^{-f/2+\lam f+\Delta u}
  \end{split}
\end{equation}
with only one independent $\Delta u$. The two equations~(\ref{eq:mod:exit})
then lead to quadratic equations for the unknown quantities $\Delta u(\lam)$
and $\ldt(\lam)$. The solutions are:
\begin{gather*}
  e^{\Delta u(\lam)} =
  \frac{1}{2cA(\lam)}\left[(c-1)+\sqrt{(c-1)^2+4c[A(\lam)]^2}\right], 
  \\
  \ldt(\lam) = 2c\cosh(f/2)[A(\lam)e^{\Delta u(\lam)}-1]
\end{gather*}
with $A(\lam)\equiv\cosh[f(\lam-1/2)]/\cosh(f/2)$. Again, both functions above
are symmetric through $A(1-\lam)=A(\lam)$.

Fig.~\ref{fig:walk}(a) shows the mean entropy production in the
$\lam$-ensemble of the original driven system with rates $w_{1,2}^\pm$. Around
$\lam=0$ and $\lam=1$ we observe a crossover between regimes of high entropy
production and one of low entropy production (something similar has been
observed for a driven particle moving in a periodic potential~\cite{mehl08}).
Fig.~\ref{fig:walk}(a) also shows that the biased entropy production coincides
with the typical entropy production in the mapped problem with rates $\tilde
w_{1,2}^{\pm}$ from Eq.~(\ref{eq:walk:rates}). The driving force in the mapped
system transforms as $\tilde f=f(1-2\lam)$. The barrier height is adjusted
through $\Delta u$ [see Fig.~\ref{fig:walk}(b)] and the barrier becomes
maximal for $\lam=1/2$ and vanishes for large $|\lam|$ with $\Delta
u\rightarrow\beta J/2$ to allow for the maximal current. One should note that
the correct expression to compare $-\ldt'(\lam)$ with is $f\mean{x}$ [where
the brackets are now the average over the dynamics with
rates~(\ref{eq:walk:rates})] and not the actual entropy production $\tilde
f\mean{x}$ of the mapped system.

%% ==== conclusions ====

\textit{Conclusions.--} We have shown that the first-order space-time phase
transitions between active and inactive phases observed in the equilibrium
dynamics of model glasses~\cite{garr07} are also present in systems driven to
non-equilibrium stationary states.  In this case the transition is between
phases with differing entropy production rates. A next step is to search for
evidence of such phase transitions in driven atomistic liquids in analogy with
what was done in Ref.~\cite{hedg09} for the case of equilibrium fluids.

%% ==== acknowledgements ====

We are grateful to David Chandler for important discussions and Lester Hedges
for help with simulating the (2)-TLG. TS acknowledges financial support by the
Alexander-von-Humboldt foundation and the Helios Solar Energy Research Center
which is supported by the Director, Office of Science, Office of Basic Energy
Sciences of the U.S. Department of Energy under Contract
No.~DE-AC02-05CH11231.

%% ==== bibliography ====


\begin{thebibliography}{99}
\bibitem{how}
This can be realised, for example, by coupling externally forced probe particles to the FA model as in Ref.\ \cite{jack08}. The back reaction on the FA model gives rise to an effective hopping of the excitations in the direction opposite to the force.  The model we introduce can be thought of as one where the probe particles have been integrated out.
\end{thebibliography}
\end{document}